\documentclass[prl,twocolumn,showpacs,amsmath,amssymb,superscriptaddress]{revtex4-1}
\usepackage{graphicx}
\usepackage{dcolumn}
\usepackage{bm}
\usepackage{bbm}
\usepackage{ulem}
\usepackage{color}
\usepackage{slashed}
\usepackage{hyperref}
\usepackage{mathrsfs}
\usepackage{subfigure}
\usepackage{verbatim}
\usepackage{dsfont}
\usepackage{float}

\newcommand{\nc}{\newcommand}

\nc{\be}{\begin{equation}} \nc{\ee}{\end{equation}}
\nc{\bea}{\begin{eqnarray}} \nc{\eea}{\end{eqnarray}}
\nc{\bean}{\begin{eqnarray*}} \nc{\eean}{\end{eqnarray*}}
\nc{\dg}{\dagger}
\nc{\ua}{\uparrow} \nc{\da}{\downarrow}

\begin{document}

\bibliographystyle{apsrev4-1}

\title{Topological nematic phase in Dirac semi-metals}
\author{Rui-Xing Zhang}
\affiliation{Department of Physics, The Pennsylvania State University, University Park, Pennsylvania 16802}
\author{Jimmy A. Hutasoit}
\affiliation{Instituut-Lorentz, Universiteit Leiden, P.O. Box 9506, 2300 RA Leiden, The Netherlands}
\author{Yan Sun}
\affiliation{Max Planck Institute for Chemical Physics of Solids, 01187 Dresden, Germany}
\author{Binghai Yan}
\affiliation{Max Planck Institute for Chemical Physics of Solids, 01187 Dresden, Germany}
\affiliation{Max Planck Institute for the Physics of Complex Systems, 01187 Dresden,Germany}
\author{Cenke Xu}
\affiliation{Department of physics, University of California, Santa Barbara, CA 93106, USA}
\author{Chao-Xing Liu}
\affiliation{Department of Physics, The Pennsylvania State University, University Park, Pennsylvania 16802}

\date{\today}

\begin{abstract}
We study the interaction effect in a three dimensional Dirac semimetal and find that two competing orders, charge-density-wave orders and nematic orders, can be induced to gap the Dirac points. Applying a magnetic field can further induce an instability towards forming these ordered phases. The charge density wave phase is similar as that of a Weyl semimetal while the nematic phase is unique for Dirac semimetals. Gapless zero modes are found in the vortex core formed by nematic order parameters, indicating the topological nature of nematic phases. The nematic phase can be observed experimentally using scanning tunnelling microscopy.
\end{abstract}


\maketitle

Dirac semimetals are materials whose bulk valence and conduction bands touch only at certain discrete momenta, around which the low energy physics is described by gapless Dirac fermions with linear energy dispersion. The two-dimensional Dirac semimetal is realized in graphene and has been studied extensively. The three-dimensional Dirac semimetals were predicted to exist in Na$_3$Bi and Cd$_3$As$_2$ \cite{young2012,wang2012,wang2013a} and these predictions were confirmed in the recent angular resolved photon emission experiments \cite{Liu2014a,Liu2014b}. The three-dimensional Dirac semimetal contains multiple copies of Weyl fermions and thus, they can exhibit non-trivial topology. Different from Weyl semimetals, the gapless nature of Dirac semimetals requires the protection of crystalline symmetries. As a consequence, by breaking some of these symmetries, Dirac semimetals can be driven towards other exotic states such as Weyl semimetals \cite{burkov2011,halasz2012,zhang2015a,lv2015,xu2015} and axionic insulators \cite{wang2013b,roy2014}.

In this letter, we consider the mass generation of a three dimensional Dirac semimetal with two Dirac points protected by rotational symmetry, such as the one realized in Na$_3$Bi. Three different complex mass terms will arise when interaction is included in the effective Hamiltonian of a three-dimensional Dirac semimetal Na$_3$Bi. One complex mass is generated by charge density wave (CDW) that involves inter-Dirac-cone scattering and breaks translational symmetry. The other two complex masses come from nematic orders that break three-fold rotational symmetry ($C_3$) by involving excitations with different spins but within a single Dirac point. Within the mean field approximation, we map the phase diagram and find that intra-Dirac-cone interaction can spontaneously break rotational symmetry and drive the system into topological nematic phases. Electron charge distribution is identified for nematic phases, which can be directly detected by scanning tunnelling microscope (STM). We further discuss localized states in topological defects as a consequence of topological nature of nematic phases. We would like to emphasize that since gapless Dirac cones are protected by rotational symmetry, a gap opening by breaking rotation symmetry can lower the energy of a Dirac semi-metal. Thus,  the presence of nematic phases is generic in rotational-symmetry protected Dirac semimetals.

Let us start by describing our model. The low energy physics of Na$_3$Bi is well captured by the $k\cdot p$ type of Hamiltonian density $H_0(\bf k)$ around the $\Gamma$ point \cite{wang2012}
\be
H_{0}(\bf k)=
\begin{pmatrix} 
M({\bf k}) & Ak_+ & 0 & 0 \\
Ak_- & -M({\bf k}) & 0 & 0 \\
0 & 0 & M({\bf k}) & -Ak_- \\
0 & 0 & -Ak_+ & -M({\bf k})  \\
\end{pmatrix}
\label{Eq:H0}
\ee
up to the second order in $k$, where $M({\bf k})=M_0-M_1k^2_z-M_2(k^2_x+k^2_y)$. The bases here are $|s,\ua\rangle,|p_+,\ua\rangle, |s,\da\rangle, |p_-,\da\rangle$, where for a basis $|\alpha,\sigma\rangle$, $\alpha=s,p_{\pm}$ is the orbital index and $\sigma=\ua,\da$ is the spin index. The above bases are also denoted as $|\frac{1}{2}\rangle,|\frac{3}{2}\rangle, |-\frac{1}{2}\rangle, |-\frac{3}{2}\rangle$ based on the total angular momentum of each state. $M_0$, $M_1$, $M_2$ and $A$ are material dependent parameters. The part of $H_0(\bf k)$ that is proportional to the identity is not important and has been neglected. The energy dispersion is $E({\bf k})=\pm\sqrt{M^2({\bf k})+A^2k_+k_-}$ and two gapless points are located at $K_i=\left(0,0,(-1)^i\sqrt{M_0/M_1}\right)$, with $i\in\{1,2\}$. The low energy effective Hamiltonian around $K_1$ and $K_2$ can be expanded from (\ref{Eq:H0}), and it is given by $\hat{H}_0 = \sum_{\mathbf k} \Psi^{\dagger}(k) \tilde{H}_0 \Psi(k)$ in the second quantized language, where
\bea
\Psi(k)&=&(c_{k,1,s,\ua},c_{k,1,p,\ua},c_{k,1,s,\da},c_{k,1,p,\da}, \nonumber \\
&& \ \ c_{k,2,s,\ua},c_{k,2,p,\ua},c_{k,2,s,\da},c_{k,2,p,\da})^T, \nonumber \\
\tilde{H}_0&=&Ak_x\alpha_0\otimes\Gamma_3-Ak_y\alpha_0\otimes\Gamma_4+m(k_z)\alpha_3\otimes\Gamma_5,
\eea
${\bf k}=(k_x,k_y,k_z)$ is the momentum relative to the Dirac points $K_i$, $m(k_z)=-2\sqrt{M_0M_1}k_z$ and $c^{\dg}_{k,i,a,\sigma}$ creates an electron with $a$ orbital and spin $\sigma$ at $K_i+k$. We also denote $c_{k,i,p_{\pm},\sigma}$ as $c_{k,i,p,\sigma}$ for brevity. $\vec{\sigma}, \vec{\tau}, \vec{\alpha}$ are Pauli matrices characterizing spin, orbital and valley degree of freedoms. $\Gamma$ matrices are defined as $\Gamma_{1,2,3}=\sigma_{1,2,3}\otimes \tau_1$, $\Gamma_4=\sigma_0\otimes\tau_2$ and $\Gamma_5=\sigma_0\otimes\tau_3$. It is easy to see that they obey Clifford algebra $\{\Gamma_i,\Gamma_j\}=2\delta_{i,j}$.

We note that $\hat{H}_0$ is the minimal model for Dirac semimetals with time reversal (TR) symmetry and inversion symmetry. To describe the effective Dirac behavior of electrons near $K_i$, we keep only the linear terms in $k$. It should be emphasized that including other higher order off-diagonal terms cannot open a gap at $K_1$ and $K_2$ since two degenerate states transform differently under three-fold rotational symmetry.

The fermionic field operator $\Psi$ can be thought of as four copies of Weyl fermions, two with left-handed chiralities and the other two right-handed. Here, we focus on the case with charge conservation and thus, the mass terms can only be formed by interactions of two Weyl fermions with opposite chiralities and therefore, there are two possible mass terms. The first one comes from two Weyl fermions with opposite chiralities at different momenta ($K_1$ and $K_2$). This term breaks translational symmetry and corresponds to CDWs. Such a term can also be found in Weyl semimetals and is responsible for axion insulator phases \cite{yang2011,wang2013b,roy2014}. Since Dirac semimetals can be viewed as two copies of Weyl semimetals that conserve TR symmetry, the gapped phase due to CDWs should also be thought of as two copies of axion insulator phases which are related to each other by TR symmetry.
The second mass term couples two Weyl fermions at the same momentum ($K_1$ or $K_2$). Since the gapless nature of Dirac semimetals at a fixed momentum is protected by $C_3$ symmetry, it is natural to expect this mass term to break rotation symmetry but preserves translational symmetry. This corresponds to a nematic phase. Therefore, these mass terms should be generated by the following order parameters:
\begin{eqnarray}
\text{CDW}&:&D_{\alpha,\beta,\sigma}=<c^{\dagger}_{k,1,\alpha,\sigma}c_{k,2,\beta,\sigma}>, \nonumber \\
\text{nematic}&:&N_{\alpha,\beta,K_i}=<c^{\dagger}_{k,i,\alpha,\uparrow}c_{k,i,\beta,\downarrow}>.
\label{Eq:Order parameter}
\end{eqnarray}
On the other hand, possible mass terms should then anti-commute with $\tilde{H}_0$ and there are only six of such terms:
$\alpha_0\otimes\Gamma_1,\ \alpha_0\otimes\Gamma_2,\ \alpha_1\otimes\Gamma_5,\alpha_2\otimes\Gamma_5,\ \alpha_3\otimes\Gamma_1,\ \alpha_3\otimes\Gamma_2.$ Based on the above analysis, we identify all possible mass terms and introduce
\bea
N^*_{s,p,1}&=&N^*_{p,s,1}=\Delta_1+\Delta_2, \nonumber \\
N^*_{s,p,2}&=&N^*_{p,s,2}=\Delta_1-\Delta_2, \nonumber \\
D^*_{s,s,\ua}&=&D^*_{s,s,\da}=-D^*_{p,p,\ua}=-D^*_{p,p,\da}=\Delta_3,
\label{Eq:massive orders}
\eea
where $\Delta_j$'s are generally complex: $\Delta_j=|\Delta_j|e^{i\theta_j}$ ($j\in1,2,3$).

To dynamically generate these mass terms, we consider an effective interaction between different species of Dirac fermions as given by
\bea
\hat{H}_{int}&=&U\sum_{k}\sum_{i}\rho_i(k)\rho_i(k)+V\sum_{k}\sum_{i\neq j}\rho_i(k)\rho_j(k),
\label{Eq:H_int}
\eea
where $\rho_i=\sum_{\alpha,\sigma}c^{\dagger}_{k,i,\alpha,\sigma}c_{k,i,\alpha,\sigma}$ are the density operators.
Here, the $U$ term describes the interaction between two electrons near one momentum $K_i$ while $V$ term describes that of electrons between $K_1$ and $K_2$. This effective interaction can be obtained from the Coulomb interaction, as shown in the Supplementary Materials \cite{supplementary}.


\begin{figure}[t]
\includegraphics[width=3.5in]{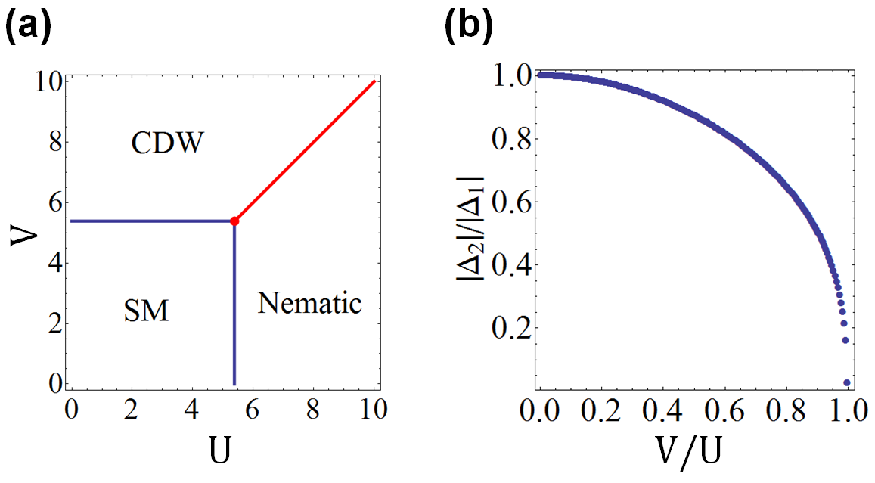}
\caption{(a) Phase diagram of interacting 3D Dirac semimetal Na$_3$Bi. (b) In the nematic phase, the ratio between $\Delta_2$ and $\Delta_1$ is plotted as a function of $V/U$. }
\label{Figure:Phase Diagram}
\end{figure}

The full Hamiltonian can then be treated within the mean field approximation (see the Supplementary Materials \cite{supplementary} for details), the free energy at zero temperature is then given by
\be
F=H_{MF}-\sum_{E_k \in {\rm occupied}}E_k (|\Delta_1|,|\Delta_2|,|\Delta_3|,\theta),
\label{Eq:free energy}
\ee
Here $E_k$ is the excitation spectrum in the mean field level, whose detailed expression is shown in the Supplementary Materials \cite{supplementary}. $\theta=\theta_1-\theta_2$ represents the phase difference between the two nematic order parameters. To minimize the free energy, a state where $\theta=\frac{\pi}{2}$ is favored. We establish self-consistency equations to map the phase diagram in Fig. \ref{Figure:Phase Diagram}(a). The semimetallic phase is relatively stable under weak interaction because the density of states vanishes at Dirac points. As the interaction strength exceeds critical value $U_c$ ($V_c$), the system develops a gap. In the large $U$ ($V$) limit, the system favors nematic (CDW) ordering. Starting from the bi-critical point $(U_c,V_c)$, the system will go across a first-order phase transition at the $U=V$ line between the CDW and nematic phases. 

The ordered phase of CDW is similar to that in Weyl semimetals, the physical consequence of which has been discussed in details in \cite{yang2011,wang2013b,roy2014}. For our system, the CDW is along the $k_z$ direction with the wave vector ${\bf Q}=2\sqrt{M_0/M_1}\hat{z}$, which can be in principle observed in an STM. Chiral modes have been proposed to exist at the dislocation line of Weyl semimetals, but since our TR invariant system is a copy of two Weyl semimetals, we have two copies of chiral modes that are TR partners and thus, our system exhibits helical modes.

What is really unique in the Dirac semimetals is the nematic phase. This nematic phase is actually a superposition of two inequivalent nematic orders $\Delta_1$ and $\Delta_2$ with a phase difference of $\frac{\pi}{2}$. By applying TR operation $\Theta=\alpha_1\otimes i\sigma_2\otimes\tau_0$, we find $\Delta_1$ breaks TR symmetry while $\Delta_2$ preserves TR symmetry. In Fig. \ref{Figure:Phase Diagram} (b), the co-existence of two nematic orders is numerically confirmed. As the ratio $V/U$ increases from 0 to 1, we find that the ratio $\Delta_2/\Delta_1$ decreases from 1 to 0. This indicates that the system spontaneously breaks TR symmetry in the nematic phase. Next, we will discuss several physical phenomena of nematic phases, which can be observed in experiments.

\begin{figure}[t]
\includegraphics[width=3.5in]{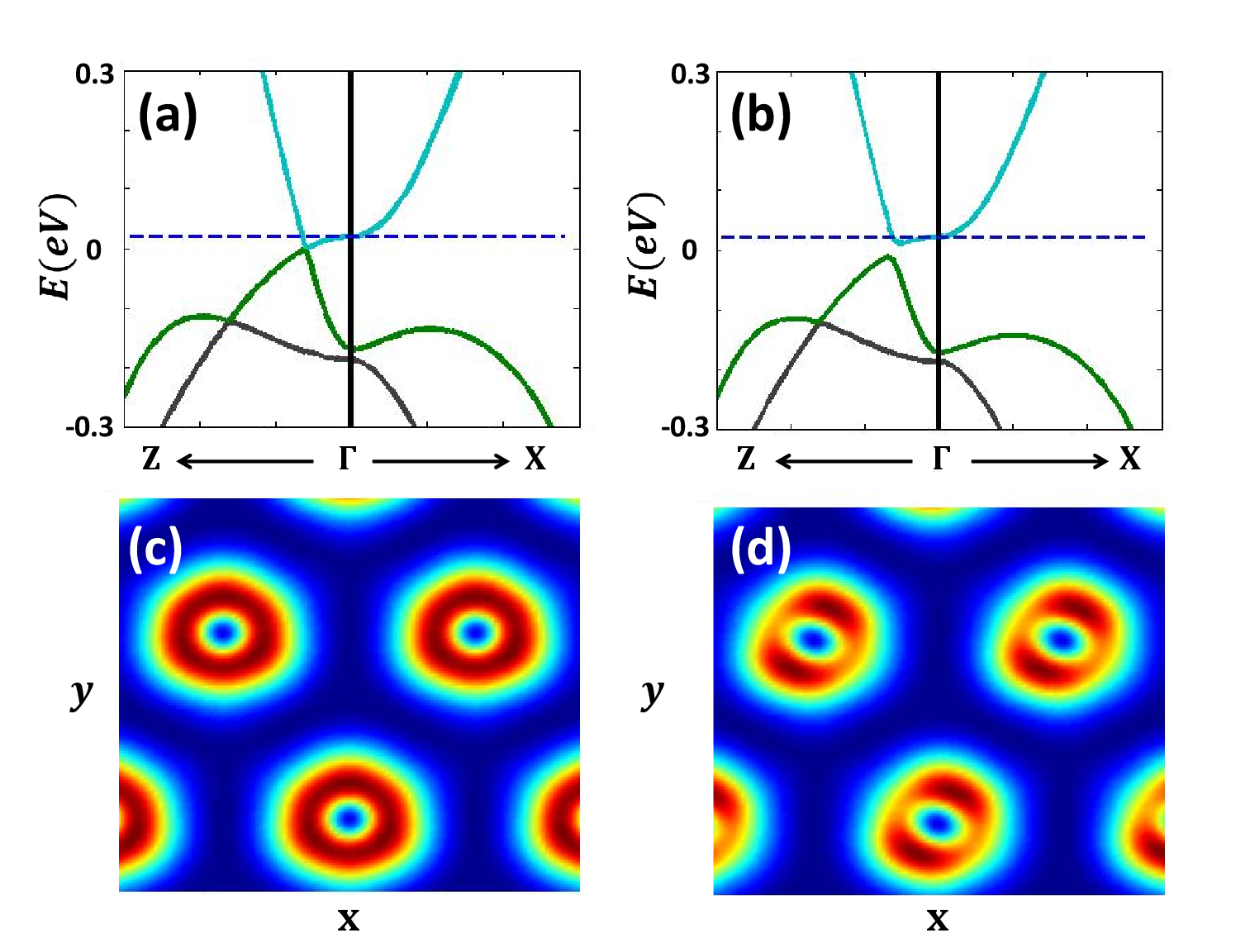}
\caption{The energy dispersion from realistic $k \cdot p$ theory and LDOS for one Bi layer. (a) and (c) are for free semimetal, while (b) and (d) are for the interacting case. In the LDOS plot, red (blue) color represent a large (small) LDOS.}
\label{Figure:LDOS}
\end{figure}

The first observable is the charge distribution. Since the mass term of nematic orders couples $|\pm\frac{3}{2}\rangle$ to $|\pm\frac{1}{2}\rangle$, we expect the charge distribution in one unit-cell breaking three fold rotation. Since the charge distribution cannot be extracted from the effective Hamiltonian, we consider a more realistic $k\cdot p$ Hamiltonian based on the first principles calculations. The method has been successfully applied to the construction of the effective Hamiltonian of topological insulator materials \cite{liu2010}, and we only describe our procedure briefly here. The eigen wave functions at $k=0$ can be expanded in term of plane waves in the first principles calculations. Here, 36 bands are taken into account, denoted as $|n\rangle$ ($n=1,2,\cdots,36$). We act the crystal Hamiltonian with periodic potential on the basis and obtain the $k\cdot p$ Hamiltonian $H^{kp}_{nm}=(E_n+\frac{\hbar^2k^2}{2m})\delta_{nm}+\frac{\hbar}{m}k\cdot p_{nm}$, where $E_n$ is the eigen-energy for the $n$ band at $k=0$, $m$ is electron mass and $p_{nm}=\langle n|p|m\rangle$ is the matrix element.
We diagonalize this $36\times36$ Hamiltonian and the energy dispersion is shown in Fig. \ref{Figure:LDOS}(a), which qualitatively fits to that from the first principles calculations. In particular, a level crossing between conduction and valence bands, which corresponds to Dirac points, can be seen along the $\Gamma-Z$ line. From the eigen wave functions, one can show that the conduction and valence bands indeed belong to the $|\pm\frac{3}{2}\rangle$ and $|\pm\frac{1}{2}\rangle$ states, respectively. Thus, these two states cannot be coupled to each other along the $\Gamma-Z$ line. As discussed above, the interaction can introduce the coupling between these two states in the nematic phase. Therefore, we can add a constant coupling between the $|\pm\frac{1}{2}\rangle$ and $|\mp\frac{3}{2}\rangle$ states near the Fermi energy in our $k\cdot p$ Hamiltonian, leading to a gap opening, as shown in Fig. \ref{Figure:LDOS}(b). To show that the obtained states possess nematic orders, we calculate the local density of states (LDOS) in the $x$-$y$ plane for the Bi layer. As shown in Fig. \ref{Figure:LDOS}(c), without interaction, the maxima of the LDOS (red color) appear as an isotropic ring around the position of Bi atoms, corresponding to the $p_\pm$ orbitals of Bi atoms. The spatial distribution of LDOS respects three-fold rotation symmetry. After adding the coupling term between $\pm|\frac{1}{2}\rangle$ and $\mp|\frac{3}{2}\rangle$ states, the isotropic ring evolves into two peaks pointing a certain direction, thus breaking $C_3$ rotation (see Fig. \ref{Figure:LDOS}(d)). This corresponds exactly to the nematic phase. Such electron density distribution can be directly measured through STM.

The second phenomenon is the appearance of gapless modes in topological defects of the nematic phase, revealing the topological nature of this phase.
Complex mass terms $\Delta=|\Delta|e^{i\theta}$ in a Dirac system are highly non-trivial in the sense that their phases $\theta$ are identified as dynamical axion fields, which will give rise to bulk axionic terms in the form of $\theta \epsilon^{\mu \nu \rho \sigma} F_{\mu \nu} F_{\rho \sigma}$ \cite{wang2013b,zyuzin2012,roy2014,peccei1977,wilczek1978,weinberg1978,wilczek1987,li2010}. In 2D Dirac systems, complex mass terms will show up as a $U(1)$ or $\mathbb{Z}_n$ vortex structure in both graphene \cite{hou2007,chamon2008} and $\pi$-flux square lattice \cite{seradjeh2008,weeks2010} in the presence of interactions. As a consequence, zero modes will localize at the vortex centers carrying fractionalized charges. In 3D Weyl/Dirac systems, those zero modes extend to 1D chiral modes that go through the center of the vortices along the $z$-direction \cite{wang2013b,roy2014}. These are known as axion strings. 
As is in the case of CDW, we expect a similar physics to occur in the vortex of nematic order parameters. 

\begin{figure}[t]
\includegraphics[width=3.5in]{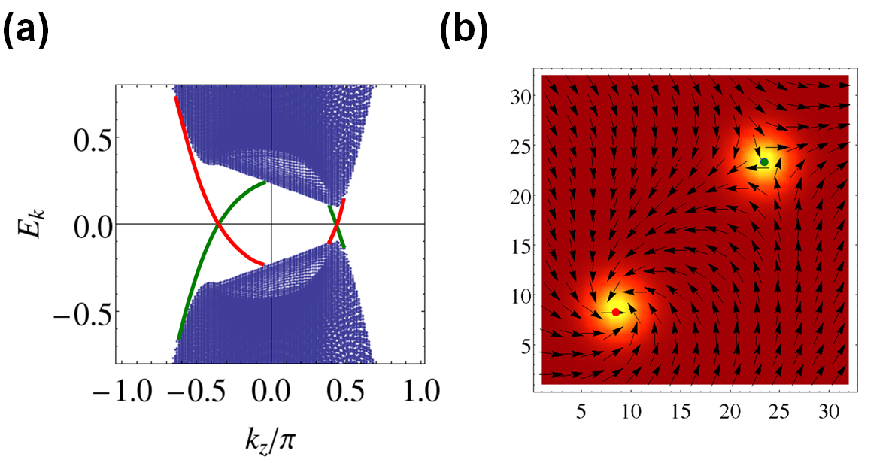}
\caption{(a) The fermionic spectrum in a $U(1)$ vortex-antivortex configuration on a $32\times32$ square lattice with open boundary conditions. We choose the following set of parameters: $M_0=-0.6, M_1=-0.3, M_2=-0.4, A=0.4, |U||\Delta_1|=0.25, |U||\Delta_2|=0.1$. Gapless energy bands in red (green) are localized at the vortex (anti-vortex) center. (b) LDOS at $E_F=0$ is plotted which clearly shows zero modes are localized at vortex or antivortex center. The red (green) dot shows the location of a vortex (antivortex) center while yellow (red) color represent a large (small) LDOS.}
\label{Figure:vortex}
\end{figure}

By applying in-plane vortex structures for the complex nematic order parameters, our system at fixed $k_z$ can be directly mapped into previous 2D Dirac systems. Therefore, zero modes are expected to show up at both $K_1$ and $K_2$. To verify this, a numerical calculation is performed in a layered 2D vortex configuration. We keep the periodicity in the $z$ direction, while placing open boundary conditions in the $x$-$y$ plane. For simplicity, on a $32\times32$ square lattice, we place a $U(1)$ vortex-antivortex pair structure instead of the actual $\mathbb{Z}_3$ vortices. We visualize these vortex structures in Fig. \ref{Figure:vortex}(b)  by the arrow indicating phase information of following site-dependent order parameters \cite{seradjeh2008}:
\bea
\tilde\Delta_1(x,y,k_z)&=&|\Delta_1|\frac{(\omega-\omega_1)(\omega-\omega_2)^*}{|(\omega-\omega_1)(\omega-\omega_2)|}, \nonumber \\
\tilde\Delta_2(x,y,k_z)&=&k_z\frac{|\Delta_2|}{|\Delta_1|}\tilde\Delta_1(x,y,k_z).
\eea
Here, $\omega=x+iy$ is a complex coordinate and $\omega_j=x_j+iy_j$ is the complex coordinate of vortex $(j=1)$ or anti-vortex center $(j=2)$. As shown in Fig. \ref{Figure:vortex}(a), the bulk dispersion is gapped while gapless modes penetrate the bulk gap twice at two different momenta.
In Fig. \ref{Figure:vortex}(b), we plot the LDOS at $E_F=0$ together with vortex configurations in real space. It is confirmed that these modes are highly localized at the vortex (anti-vortex) center. The gapless nature of these modes relies on the fact that they are separated at different momenta, and requires the translational symmetry along the z-direction.

\begin{figure}[t]
\includegraphics[width=3.5in]{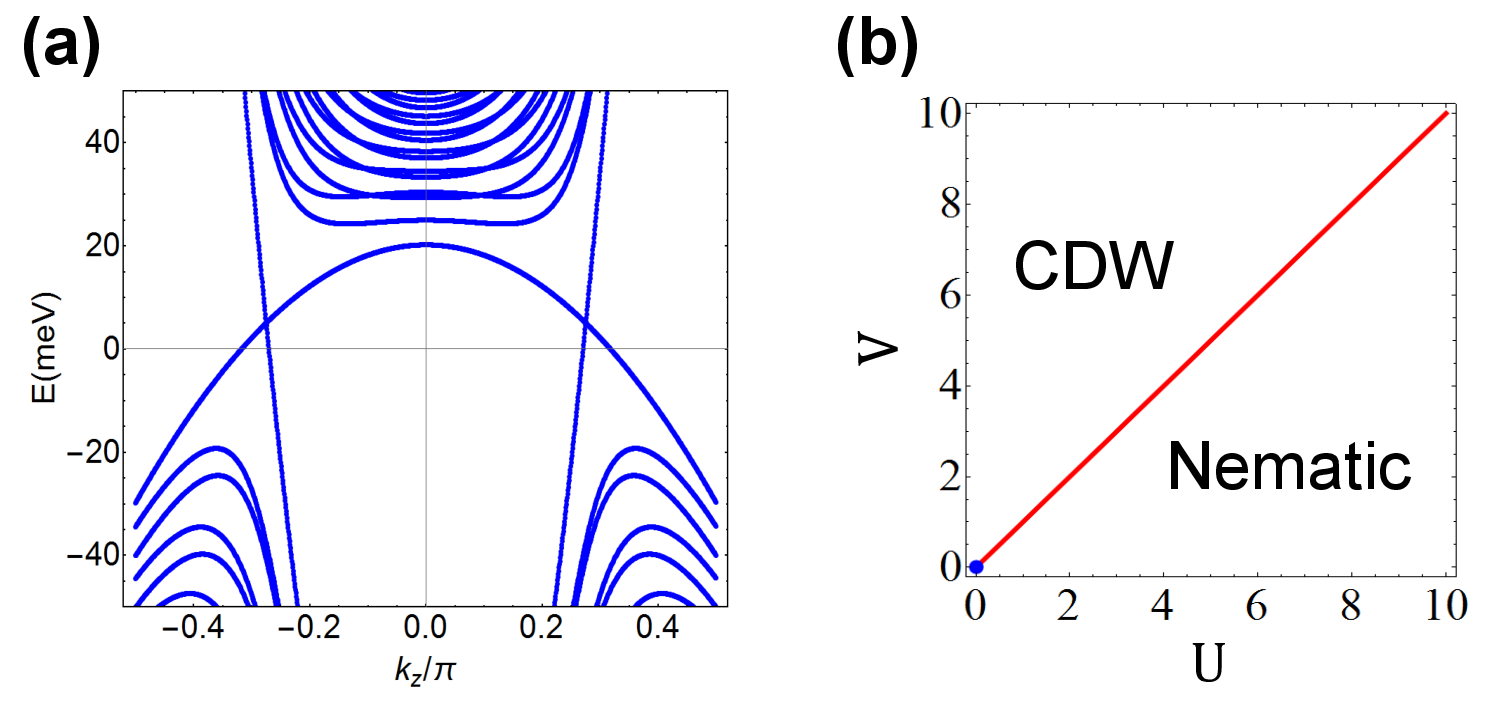}
\caption{(a) Landau level dispersion along $k_z$ with magnetic field $B=10$T. (b) Phase diagram of Na$_3$Bi under magnetic field and interaction.}
\label{Figure:Landau Level}
\end{figure}

So far, we have discussed the effects of interaction in driving Dirac semimetals toward other phases. However, those phases along with their novel physical phenomena can only be realized under relatively strong interaction. To overcome this difficulty, one can apply a magnetic field along the $z$ direction such that Landau levels emerge. Similar strategies have been applied to achieve quantum Hall ferromagnetism in graphene systems \cite{nomura2006,young2012b}, where spin orbital coupling (SOC) is almost absent. The strong SOC in Dirac semimetals, however, tends to tilt spins. As a result, CDW and nematic phases are more likely to be favored than ferromagnetism in Dirac semimetals.
The Landau levels in Dirac semi-metals have been observed experimentally \cite{jeon2014,he2014,zhao2014,kushwaha2015,zhao2015}.
Even though the higher Landau levels of Na$_3$Bi are gapped, the lowest Landau levels (LLLs) are gapless at $K_i$ ($i=1,2$), see Fig. \ref{Figure:Landau Level}. We identify this degeneracy to be a crossing between $|s,\ua\rangle$ and $|p,\da\rangle$ states, which is protected from developing a gap by $C_3$ symmetry along (001) axis. To describe the low energy physics of the gapless LLLs, we define a four-component spinor, $\Psi^{\dagger}=(c^{\dagger}_{k,1,s,\uparrow},c^{\dagger}_{k,1,p,\downarrow},c^{\dagger}_{k,2,s,\uparrow},c^{\dagger}_{k,2,p,\downarrow})$. Mass terms in Eq. \eqref{Eq:massive orders} are reduced to: (1) Density Wave: $D_1=D_{s,s,\uparrow},\ D_2=D_{p,p,\downarrow}$. (2) Nematic: $N_1=N_{s,p,K_1},\ N_2=N_{s,p,K_2}$. Through a similar mean field analysis (see the Supplementary Materials), the free energy at zero temperature is given by $F=H_{MF}-\sum_{i}\sum_{k_z}\sqrt{m(k_z)^2+\xi_i}$
where $m(k_z)=-2\sqrt{M_1(M_0-\frac{M_2}{l^2})}k_z$. $\xi_{1,2}$ are functions of order parameters $D_{1,2}$ and $N_{1,2}$, whose detailed expressions are explicitly shown in the Supplementary Materials \cite{supplementary}.

By minimizing the free energy, we obtain the phase diagram as shown in Fig. \ref{Figure:Landau Level}(b). Instability happens for arbitrarily weak repulsive interaction \cite{yang2011, roy2014} and as one tunes the interaction to go across $V/U=1$, the system undergoes a phase transition from a CDW phase to a nematic phase or vice versa. 
Let us focus on the nematic regime ($D_{1,2}=0$) and the corresponding self-consistent equations can be solved analytically. As is shown in the Supplementary Materials \cite{supplementary}, the critical temperature that characterizes a phase transition from semi-metallic phase to the nematic phase is
\be
T_c=\frac{2e^{\gamma}v_f{\Lambda}}{\pi k_B}e^{-\frac{v_f\Lambda}{U}\frac{h}{eBS}}
\label{Eq:Tc}
\ee
where $\Lambda$ is the momentum cut-off and $v_f=|\frac{m(k_z)}{k_z}|$ is the Fermi velocity. $\gamma=0.577...$ is the Euler constant and $k_B$ is the Boltzmann constant. We have considered a sample with a finite area $S$ in the $x$-$y$ plane. When $T<T_c$, non-zero nematic ordering will always be formed for arbitrary $U$. In the zero temperature limit, the magnitude of order parameter can be solved \cite{supplementary,shankar1994}: $|N_1|\approx\frac{2v_f\Lambda}{U}e^{-\frac{2\pi v_f\Lambda}{U}\frac{h}{eBS}}$. This expression indicates that a larger energy gap will show up for a larger magnetic field. This instability under magnetic fields is a direct result of the finite Landau level degeneracy. This suggests the necessary condition for the instability is that the cyclotron length is much smaller than the sample size. In the Supplementary Materials \cite{supplementary}, we further discuss the existing experiments studying LLs of Dirac semimetals, and predict possible evidence of nematic phases in STM measurements of Landau levels. 

\textit{Acknowledgement -} We acknowledge the helpful discussion with XL Qi. C.X.L is supported by ONR (Grant No. N00014-15-1-2675). J. H. is supported by NSF grant DMR-1005536 and DMR-0820404 (Penn State MRSEC) during the early part of this work, and later by the Netherlands Organization for Scientific Research (NWO/OCW) through the D-ITP consortium. Cenke Xu is supported by the the David and Lucile Packard Foundation and NSF Grant No. DMR-1151208.

\bibliography{na3bi}

\onecolumngrid
\newpage

\subsection{\large Supplementary Materials for ``Nematic phase of Dirac semimetal"}

\subsection{Microscopic derivation of $H_{int}$ from Coulomb interaction}
In this section, we give a microscopic derivation of the interacting term $H_{int}$ from the well-known Coulomb interaction,
\be
H_{Coulomb}=\sum_{k,k',q}\sum_{\alpha,\beta,\sigma,\sigma'}V(q)c^{\dagger}_{k+q,\alpha,\sigma}c_{k,\alpha,\sigma}c^{\dagger}_{k'-q,\beta,\sigma'}c_{k',\beta,\sigma'}.
\ee
Here, $V(q)=2\pi e^2/|q|^2$. $\alpha$ and $\beta$ are orbital indices, while $\sigma$ and $\sigma'$ are spin indices. Then we apply a mean field treatment to $H_{Coulomb}$,
\bea
H_{Coulomb}&=&\sum_{k,k',q}\sum_{\alpha,\beta,\sigma,\sigma'}V(q)c^{\dagger}_{k+q,\alpha,\sigma}c_{k,\alpha,\sigma}
c^{\dagger}_{k'-q,\beta,\sigma'}c_{k',\beta,\sigma'} \nonumber \\
&=&-\sum_{k,k',q}\sum_{\alpha,\beta,\sigma,\sigma'}V(q)[c^{\dagger}_{k+q,\alpha,\sigma}c_{k',\beta,\sigma'}-\langle c^{\dagger}_{k+q,\alpha,\sigma}c_{k',\beta,\sigma'} \rangle+\langle c^{\dagger}_{k+q,\alpha,\sigma}c_{k',\beta,\sigma'} \rangle] \nonumber \\
&&\times [c^{\dagger}_{k'-q,\beta,\sigma'}c_{k,\alpha,\sigma}-\langle c^{\dagger}_{k'-q,\beta,\sigma'}c_{k,\alpha,\sigma} \rangle+ \langle c^{\dagger}_{k'-q,\beta,\sigma'}c_{k,\alpha,\sigma} \rangle] \nonumber \\
&\approx& \sum_{k,k',q}\sum_{\alpha,\beta,\sigma,\sigma'}V(q) [\langle c^{\dagger}_{k+q,\alpha,\sigma}c_{k',\beta,\sigma'} \rangle \times \langle c^{\dagger}_{k'-q,\beta,\sigma'}c_{k,\alpha,\sigma} \rangle \nonumber \\
&&-\langle c^{\dagger}_{k'-q,\beta,\sigma'}c_{k,\alpha,\sigma} \rangle c^{\dagger}_{k+q,\alpha,\sigma}c_{k',\beta,\sigma'}
-\langle c^{\dagger}_{k+q,\alpha,\sigma}c_{k',\beta,\sigma'} \rangle c^{\dagger}_{k'-q,\beta,\sigma'}c_{k,\alpha,\sigma}] .\nonumber \\
\eea
Naively, we are particularly interested in the scattering process between Weyl fermions with opposite chirality. As shown in the main text, we have identified all possible mass terms (order parameters):
\begin{eqnarray}
\text{CDW}&:& <c^{\dagger}_{k+q,\alpha,\sigma}c_{k',\beta,\sigma}>=D_{\alpha,\beta,\sigma}\times\delta_{k,k'-q-2K_0}, \nonumber \\
\text{Nematic}&:& <c^{\dagger}_{k+q,\alpha,\uparrow}c_{k',\beta,\downarrow}>=N_{\alpha,\beta,K_i}\times\delta_{k,k'-q}.
\label{Eq:order_coulomb}
\end{eqnarray}
In the definition of nematic order $N_{\alpha,\beta,K_i}$, we have defined that both $k$ and $k'$ are effective momenta relative to bulk Dirac point $K_{i=1,2}$. Based on Eq. \eqref{Eq:order_coulomb}, we are ready to decompose $H_{Coulomb}$ into different channels $H_{Coulomb}=H_{CDW}+H_{Nematic}$:
\bea
H_{CDW}&=&\sum_{k,k',q}\sum_{\alpha,\beta,\sigma}V(q) (|D_{\alpha,\beta,\sigma}|^2 \times\delta_{k,k'-q-2K_0} -D^*_{\alpha,\beta,\sigma}\times\delta_{k,k'-q-2K_0} c^{\dagger}_{k+q,\alpha,\sigma}c_{k',\beta,\sigma} \nonumber \\
&&-D_{\alpha,\beta,\sigma}\times\delta_{k,k'-q-2K_0} c^{\dagger}_{k'-q,\beta,\sigma}c_{k,\alpha,\sigma}) \nonumber \\
&=& \sum_{k,q}\sum_{\alpha,\beta,\sigma}V(q) (|D_{\alpha,\beta,\sigma}|^2 -D^*_{\alpha,\beta,\sigma}\times c^{\dagger}_{k+q,\alpha,\sigma}c_{k+q+2K_0,\beta,\sigma}-D_{\alpha,\beta,\sigma}\times c^{\dagger}_{k+2K_0,\beta,\sigma}c_{k,\alpha,\sigma}) \nonumber \\
&=& \sum_{k,q}\sum_{\alpha,\beta,\sigma}V(q) (|D_{\alpha,\beta,\sigma}|^2 -D^*_{\alpha,\beta,\sigma}\times c^{\dagger}_{k,\alpha,\sigma}c_{k+2K_0,\beta,\sigma}-D_{\alpha,\beta,\sigma}\times c^{\dagger}_{k+2K_0,\beta,\sigma}c_{k,\alpha,\sigma}) \nonumber \\
&=&V \sum_{k}\sum_{\alpha,\beta,\sigma} (|D_{\alpha,\beta,\sigma}|^2 -D^*_{\alpha,\beta,\sigma}\times c^{\dagger}_{k,\alpha,\sigma}c_{k+2K_0,\beta,\sigma}-D_{\alpha,\beta,\sigma}\times c^{\dagger}_{k+2K_0,\beta,\sigma}c_{k,\alpha,\sigma}).
\label{Eq:H_CDW}
\eea
Here, $k$ is the effective crystal momenta relative to the Dirac point $(0,0,-K_0)$, therefore $k+2K_0$ is in the vicinity of the other Dirac point $(0,0,K_0)$. When discussing the nematic phase below, $k$ is the effective crystal momenta relative to the Dirac point $K_{i=1,2}=(0,0,(-1)^iK_0)$, depending on the $i$ index of fermionic operator $c_{k,i,\alpha,\sigma}$:
\bea
H_\text{Nematic}&=&\sum_{i=1,2}\sum_{k,k',q}\sum_{\alpha,\beta}V(q) [|N_{\alpha,\beta,i}|^2 \times\delta_{k,k'-q} -N^*_{\alpha,\beta,i}\times\delta_{k,k'-q} c^{\dagger}_{k+q,i,\alpha,\ua}c_{k',i,\beta,\da} \nonumber \\
&&-N_{\alpha,\beta,i}\times\delta_{k,k'-q} c^{\dagger}_{k'-q,i,\beta,\da}c_{k,i,\alpha,\ua}] \nonumber \\
&=&\sum_{i=1,2}\sum_{k,q}\sum_{\alpha,\beta}V(q) [|N_{\alpha,\beta,i}|^2 -N^*_{\alpha,\beta,i}\times c^{\dagger}_{k+q,i,\alpha,\ua}c_{k+q,i,\beta,\da}-N_{\alpha,\beta,i}\times c^{\dagger}_{k,i,\beta,\da}c_{k,i,\alpha,\ua}] \nonumber \\
&=&\sum_{i=1,2}\sum_{k,q}\sum_{\alpha,\beta}V(q) [|N_{\alpha,\beta,i}|^2 -N^*_{\alpha,\beta,i}\times c^{\dagger}_{k,i,\alpha,\ua}c_{k,i,\beta,\da}-N_{\alpha,\beta,i}\times c^{\dagger}_{k,i,\beta,\da}c_{k,i,\alpha,\ua}] \nonumber \\
&=&U\sum_{i=1,2}\sum_{k}\sum_{\alpha,\beta} [|N_{\alpha,\beta,i}|^2 -N^*_{\alpha,\beta,i}\times c^{\dagger}_{k,i,\alpha,\ua}c_{k,i,\beta,\da}-N_{\alpha,\beta,i}\times c^{\dagger}_{k,i,\beta,\da}c_{k,i,\alpha,\ua}].
\label{Eq:H_nem}
\eea
In the above expressions, we have defined a CDW (Nematic) interaction strength $U$ ($V$). It is interesting to notice that $U$ and $V$ take the same value $\sum_{q}2\pi e^2/|q|^2$. In our phase diagram of mean field theory, $U=V$ corresponds to the critical line separating CDW phase with nematic phase. However, in realistic materials, we expect one of the two phases will be favored, depending on the material details, which is beyond the scope of this paper.

\section{Mean Field Theory}
Starting from Coulomb interaction, we have shown that the essential physics is captured by inter-Dirac-cone scattering ($H_{CDW}$) and intra-valley-scattering ($H_\text{Nematic}$). This inspires us to write the effective density-density interaction Eq. (5) in the main text:
\bea
\hat{H}_{int}&=&U\sum_{k}\sum_{i}\rho_i(k)\rho_i(k)+V\sum_{k}\sum_{i\neq j}\rho_i(k)\rho_j(k),
\eea
where $\rho_i(k)=\sum_{\alpha,\sigma}c^{\dagger}_{k,i,\alpha,\sigma}c_{k,i,\alpha,\sigma}$ are the density operators. This effective interaction term is equivalent to both Eq. \eqref{Eq:H_CDW} and Eq. \eqref{Eq:H_nem}, while illustrating the physics in a better way. Based on the form of order parameters, we could put constraints to the indices and further simplify the density-density interaction to be
\begin{eqnarray}
\rho_i\rho_i&=&\sum_k \sum_{\alpha,\beta,\sigma\neq\sigma'}c^{\dagger}_{k,i,\alpha,\sigma}c_{k,i,\alpha,\sigma}
c^{\dagger}_{k,i,\beta,\sigma'}c_{k,i,\beta,\sigma'}, \nonumber \\
\rho_i\rho_j&=&\sum_k \sum_{\alpha,\beta,\sigma}c^{\dagger}_{k,i,\alpha,\sigma}c_{k,i,\alpha,\sigma}
c^{\dagger}_{k,j,\beta,\sigma}c_{k,j,\beta,\sigma}.
\end{eqnarray}
Applying a similar mean field analysis to our earlier discussion, the interaction terms can then be written as
\begin{eqnarray}
\rho_i\rho_i&=&\sum_k\sum_{\alpha,\beta}(|N_{\alpha,\beta,K_i}|^2-N_{\alpha,\beta,K_i}c^{\dagger}_{k,i,\beta,\downarrow}c_{k,i,\alpha,\uparrow}
-N^{*}_{\alpha,\beta,K_i}c^{\dagger}_{k,i,\alpha,\uparrow}c_{k,i,\beta,\downarrow}), \nonumber \\
\rho_i\rho_j&=&\sum_k\sum_{\alpha,\beta,\sigma}(|D_{\alpha,\beta,\sigma}|^2-D_{\alpha,\beta,\sigma}c^{\dagger}_{k,2,\beta,\sigma}c_{k,1,\alpha,\sigma}
-D^{*}_{\alpha,\beta,\sigma}c^{\dagger}_{k,1,\alpha,\sigma}c_{k,2,\beta,\sigma}),
\end{eqnarray}
where the order parameters are defined in Eq. (3) of the main article.


Then, the mean field Hamiltonian is readily obtained
\bea
H&=&\sum_{k} \Psi^{\dagger}H_{int} \Psi+H_{MF}, \nonumber \\
H_{int}&=&\begin{pmatrix} 
H_{11} & H_{12} \\
H_{12}^{\dg} & H_{22} \\
\end{pmatrix}, \nonumber \\
H_{MF}&=&\sum_k\sum_{\alpha,\beta=s,p}U(|N_{\alpha,\beta,1}|^2+|N_{\alpha,\beta,2}|^2)+V(|D_{\alpha,\beta,\ua}|^2+|D_{\alpha,\beta,\da}|^2),
\eea
where
\be
\Psi(k)=(c_{k,1,s,\ua},c_{k,1,p,\ua},c_{k,1,s,\da},c_{k,1,p,\da},c_{k,2,s,\ua},c_{k,2,p,\ua},c_{k,2,s,\da},c_{k,2,p,\da})^T,
\ee
and $H_{int}$ is an $8\times8$ matrix with each $H_{ij}$ to be a $4\times4$ block:
\bea
H_{11}&=&
\begin{pmatrix} 
m(k) & Ak_+ & -UN^*_{s,s,1} & -UN^*_{s,p,1} \\
Ak_- & -m(k) & -UN^*_{p,s,1} & -UN^*_{p,p,1} \\
-UN_{s,s,1} & -UN_{p,s,1} & m(k) & -Ak_- \\
-UN_{s,p,1} & -UN_{p,p,1} & -Ak_+ & -m(k)  \\
\end{pmatrix}, \nonumber \\
H_{12}&=&V
\begin{pmatrix}
-D^*_{s,s,\ua} & -D^*_{s,p,\ua} & 0 & 0 \\
-D^*_{p,s,\ua} & -D^*_{p,p,\ua} & 0 & 0 \\
 0 & 0 & -D^*_{s,s,\da} & -D^*_{s,p,\da} \\
 0 & 0 & -D^*_{p,s,\da} & -D^*_{p,p,\da} \\
\end{pmatrix}.
\eea

Since we are only interested in mass terms that can gap the system, we would like to only keep mean field terms that anti-commute with the original Hamiltonian:
\bea
N^*_{s,s,i}&=&N^*_{p,p,i}=0 \nonumber \\
D^*_{s,p,\sigma}&=&D^*_{p,s,\sigma}=0 \nonumber \\
N^*_{s,p,1}&=&N^*_{p,s,1}=\Delta_1+\Delta_2 \nonumber \\
N^*_{s,p,2}&=&N^*_{p,s,2}=\Delta_1-\Delta_2 \nonumber \\
D^*_{s,s,\ua}&=&D^*_{s,s,\da}=\Delta_3 \nonumber \\
D^*_{p,p,\ua}&=&D^*_{p,p,\da}=-\Delta_3
\eea
Here, $\Delta_1$ ($\Delta_2$) is the nematic order that spontaneously breaks (preserves) TR symmetry and breaks three-fold rotational symmetry. $\Delta_{3}$ is the charge density wave order parameters that breaks translational symmetry. Also notice that these order parameters are generally complex: $\Delta_j=|\Delta_j|e^{i\theta_j}$ ($j\in1,2,3$). Then, we can write down $H_{int}$ in a compact form:
\bea
H_{int}&=&\tilde{H}_0-H_1, \nonumber \\
\tilde{H}_0&=&Ak_x\alpha_0\otimes\Gamma_3-Ak_y\alpha_0\otimes\Gamma_4+m(k)\alpha_3\otimes\Gamma_5, \nonumber \\
H_1&=&U|\Delta_1|(\cos\theta_1\alpha_0\otimes\Gamma_1-\sin\theta_1\alpha_0\otimes\Gamma_2)
+U|\Delta_2|(\cos\theta_2\alpha_3\otimes\Gamma_1-\sin\theta_2\alpha_3\otimes\Gamma_2) \nonumber \\
&&+V|\Delta_3|(\cos\theta_3\alpha_1\otimes \Gamma_5-\sin\theta_3\alpha_2\otimes \Gamma_5).
\eea

The full Hamiltonian is then given by
\bea
H&=& \sum_{\bf k} \Psi^{\dagger}(\tilde{H}_0-H_1)\Psi+H_{MF}, \nonumber \\
H_{MF}&=&4(\frac{L\Lambda}{\pi})^3[U(|\Delta_1|^2+|\Delta_2|^2)+V|\Delta_3|^2].
\eea
and $L^3$ is the volume of the sample and $\Lambda$ is the momentum cut-off. The first term can be diagonalized analytically to yield the eigen-energy
\bea
E_k&=&\pm[U^2(|\Delta_1|^2+|\Delta_2|^2)+V^2|\Delta_3|^2+A^2k_+k_-+m(k_z)^2 \nonumber \\
&&\pm2U|\Delta_2|\sqrt{V^2|\Delta_3|^2+U^2|\Delta_1|^2\cos^2(\theta_1-\theta_2)}]^{\frac{1}{2}}.
\label{Eq:Ek}
\eea
The above expression is the excitation spectrum that shows up in the free energy in the main text.

\section{Analytical Properties of Free Energy in Eq. (8) of the Main Article}
Let us first show that why $\theta=\frac{\pi}{2}$ is favored. Let us define
\bea
\epsilon({\bf k})&=&U^2(|\Delta_1|^2+|\Delta_2|^2)+V^2|\Delta_3|^2+A^2k_+k_-+m(k)^2, \nonumber \\
f({\bf k},\theta)&=&2U|\Delta_2|\sqrt{V^2|\Delta_3|^2+U^2|\Delta_1|^2\cos^2\theta},
\eea
such that the free energy can be written as
\bea
F&=&H_{MF}-2\sum_{\bf k}J({\bf k},\theta), \nonumber \\
J({\bf k},\theta)&=&\sqrt{\epsilon({\bf k})+f({\bf k},\theta)}+\sqrt{\epsilon({\bf k})-f({\bf k},\theta)}.
\eea
Notice that $H_{MF}$ is independent of $\theta$, and
\be
\frac{dJ}{df}=\frac{1}{2}[\frac{1}{\sqrt{\epsilon({\bf k})+f({\bf k},\theta)}}-\frac{1}{\sqrt{\epsilon({\bf k})-f({\bf k},\theta)}}]<0,
\ee
$f({\bf k},\theta)_{min}=f({\bf k},\theta=\frac{\pi}{2})=2UV|\Delta_2\Delta_3|$. So $f({\bf k},\theta=\frac{\pi}{2})$ will maximize J and thus minimize free energy F. So this condition constrains $\theta=\frac{\pi}{2}$.

Now we are ready to write down the self-consistency equations:
\bea
\Delta_1&=&\frac{1}{4U}\frac{1}{(2\Lambda)^3}\int d^3k\frac{\partial J}{\partial \Delta_1}, \nonumber \\
\Delta_2&=&\frac{1}{4U}\frac{1}{(2\Lambda)^3}\int d^3k\frac{\partial J}{\partial \Delta_2}, \nonumber \\
\Delta_3&=&\frac{1}{4V}\frac{1}{(2\Lambda)^3}\int d^3k\frac{\partial J}{\partial \Delta_3}.
\label{Eq:self-consistency eq}
\eea
Here $\Lambda$ is the momentum cutoff in the integration. The self-consistency equations can be solved numerically and the solution gives rise to the phase diagram in Fig. 1 of the main article. Analytically, they can also give us some hints on the shape of the phase boundary. After some manipulations, the first and the third equations in Eq. (\ref{Eq:self-consistency eq}) are:
\bea
\frac{1}{U}&=&\frac{1}{4}\frac{1}{(2\Lambda)^3}\int d^3k \frac{1}{\sqrt{\epsilon+f}}+ \frac{1}{\sqrt{\epsilon-f}}, \nonumber \\
1&=&\frac{1}{4}\frac{1}{(2\Lambda)^3}\int d^3k \frac{V+U\frac{\Delta_2}{\Delta_3}}{\sqrt{\epsilon+f}}+\frac{V-U\frac{\Delta_2}{\Delta_3}}{\sqrt{\epsilon-f}}.
\eea
By setting $\Delta_i=0$, we arrive at the critical interaction strength
\be
\frac{1}{U_c}=\frac{1}{V_c}=\frac{1}{2}\frac{1}{(2\Lambda)^3}\int d^3k\frac{1}{\sqrt{A^2k_-k_++m(k)^2}}.
\ee

\section{Landau Level of Na$_3$Bi and the self-consistent equations}

Under a magnetic field that is oriented along the $z$-direction, minimal coupling requires $\pi=k+\frac{e}{\hbar}A$. Defining the magnetic length to be $l=\sqrt{\frac{\hbar}{eB}}$, the commutation relation of $\pi$ is then given by
\begin{equation}
[\pi_x,\pi_y]=-\frac{ieB}{\hbar}=-\frac{i}{l^2},
\end{equation}
where we have chosen the gauge $A=(0,Bx,0)$. We can then define creation and annihilation operators in terms of $\pi$ as follows
\begin{eqnarray}
a=\frac{l}{\sqrt{2}}\pi_{-}, \quad
a^{\dagger}=\frac{l}{\sqrt{2}}\pi_{+}, \quad [a,a^{\dagger}]=1.
\end{eqnarray}
From the commutation relation, we find that
\begin{equation}
\pi_x^2+\pi_y^2=\frac{2}{l^2}(a^{\dagger}a+\frac{1}{2}).
\end{equation}
Then, by choosing the following trial wave-function $\Psi=(f_1^N\phi_N,f_2^N\phi_{N-1},f_3^N\phi_{N-1},f_4^N\phi_N)^{T}$, the Hamiltonian density can be written down as
\begin{eqnarray}
H(k_z,N) &=& \left(
\begin{array}{cccc}
M & A\pi_{+} & 0 & 0 \\
A\pi_{-} & -M & 0 & 0 \\
0 & 0 & M & -A\pi_{-}  \\
0 & 0 & -A\pi_{+} &-M \\
\end{array}
\right) \nonumber \\
&=& \left(
\begin{array}{cccc}
\tilde{M}_N^{+} & \frac{A}{l}\sqrt{2N} & 0 & 0 \\
\frac{A}{l}\sqrt{2N} & \tilde{M}_{N-1}^{-} & 0 & 0 \\
0 & 0 & \tilde{M}_{N-1}^{+} & -\frac{A}{l}\sqrt{2N}  \\
0 & 0 & -\frac{A}{l}\sqrt{2N} & \tilde{M}_N^{-} \\
\end{array}
\right),
\label{Eq:Landau level}
\end{eqnarray}
where
\begin{eqnarray}
&&M=M_0-M_1k_z^2-\frac{2M_2}{l^2}(a^{\dagger}a+\frac{1}{2}), \nonumber \\
&&a^{\dagger}\phi_{N-1}=\sqrt{N}\phi_N, \quad a\phi_N=\sqrt{N}\phi_{N-1}.
\end{eqnarray}

The Lowest Landau levels (LLL) are then given by $N=0$,
\begin{equation}
H(k_z,0) = \left(
\begin{array}{cccc}
\tilde{M}_0^{+} & 0 & 0 & 0 \\
0 & 0 & 0 & 0 \\
0 & 0 & 0 & 0  \\
0 & 0 & 0 & \tilde{M}_0^{-} \\
\end{array}
\right),
\end{equation}
where
\begin{equation}
\tilde{M}_0^{\pm}(k_z)=\pm M_0 \mp M_1k_z^2\mp \frac{M_2}{l^2}.
\end{equation}
This indicates that only the LLLs from $|\frac{1}{2}\rangle$ and $|-\frac{3}{2}\rangle$ states are gapless, with the gapless nodes located at $K_{i}=(0,0,(-1)^i\sqrt{\frac{1}{M_1}(M_0-\frac{M_2}{l^2})})$.

When considering instability problem of Na$_{3}$Bi under strong magnetic field, the gapless LLLs are composed of the following states:
\begin{equation}
|\frac{1}{2}\rangle=|s,\uparrow\rangle,\ \  |-\frac{3}{2}\rangle=|p,\downarrow\rangle,
\end{equation}
where we could define the following order parameters,
\begin{eqnarray}
\text{Nematic}&:& N_1=N_{s,p,K_1},\ N_2=N_{s,p,K_2} \nonumber \\
\text{Density Wave}&:& D_1=D_{s,s,\uparrow},\ D_2=D_{p,p,\downarrow}
\end{eqnarray}
The Hamiltonian is then given by
\begin{equation}
H=\sum_{k_z} \Psi^{\dagger}(H_0+H_{\rm int})\Psi + H_{\rm MF},
\end{equation}
where
\begin{eqnarray}
H_0&=&m(k_z)\tau_z\otimes\sigma_z, \nonumber \\
H_{int}&=&
\begin{pmatrix} 
0 & -UN^{*}_1 & -VD^{*}_1 & 0 \\
-UN_1 & 0 & 0 & -VD^{*}_2 \\
-VD_1 & 0 & 0 & -UN^{*}_2 \\
0 & -VD_2 & -UN_2 & 0  \\
\end{pmatrix}, \nonumber \\
H_{MF}&=&\sum_k [U(|N_1|^2+|N_2|^2)+V(|D_1|^2+|D_2|^2)].
\label{Eq:H_LL}
\end{eqnarray}
Here, $m(k_z)=-2\sqrt{M_1(M_0-\frac{M_2}{l^2})}k_z$. The matrix part $H_0+H_{int}$ can be diagonalized analytically, and the eigen-energy for occupied bands are $-\sqrt{m(k_z)^2+\xi_i}$.
\begin{eqnarray}
\xi_1&=&\frac{1}{2}[U^2(|N_1|^2+|N_2|^2)+V^2(|D_1|^2+|D_2|^2)+\sqrt{U^4(|N_1|^2-|N_2|^2)^2+V^4(|D_1|^2-|D_2|^2)^2} \nonumber \\
&&\overline{+2U^2V^2(|N_1|^2+|N_2|^2)(|D_1|^2+|D_2|^2)+8U^2V^2|N_1N_2D_1D_2|\cos[\phi_{D_1}-\phi_{D_2}-\phi_{N_1}+\phi_{N_2}]}] \nonumber \\
\xi_2&=&\frac{1}{2}[U^2(|N_1|^2+|N_2|^2)+V^2(|D_1|^2+|D_2|^2)-\sqrt{U^4(|N_1|^2-|N_2|^2)^2+V^4(|D_1|^2-|D_2|^2)^2} \nonumber \\
&&\overline{+2U^2V^2(|N_1|^2+|N_2|^2)(|D_1|^2+|D_2|^2)+8U^2V^2|N_1N_2D_1D_2|\cos[\phi_{D_1}-\phi_{D_2}-\phi_{N_1}+\phi_{N_2}]}].
\label{Eq:Ek_LL}
\end{eqnarray}

Since we are especially interested in the magnetic instability in the nematic regime, we can set density order parameters $|D_1|=|D_2|=0$. Then in the mean field level, single particle Hamiltonian $H_0+H_{\rm int}$ have four energy eigenvalues: $E_1^{\pm}=\pm\sqrt{m(k_z)^2+U^2|N_1|^2}$ and $E_2^{\pm}=\pm\sqrt{m(k_z)^2+U^2|N_2|^2}$. Quantum partition function at finite temperature $k_BT=\frac{1}{\beta}$ ($k_B$ is the Boltzmann constant) is given by
\bea
Z&=&Tr e^{-\beta H} \nonumber \\
&=&e^{-\beta H_{MF}}\text{tr} [e^{-\beta\sum_{k_z}\Psi^{\dagger}(H_0+H_{\rm int})\Psi } ] \nonumber \\
&=&e^{-\beta \sum_{k_z} U(|N_1|^2+|N_2|^2)}\times\sum_{k_z}(1+e^{-\beta E_1^{+}})(1+e^{-\beta E_1^{-}})(1+e^{-\beta E_2^{+}})(1+e^{-\beta E_2^{-}}) ] \nonumber \\
&=&e^{-\beta \frac{2L\Lambda}{2\pi} U(|N_1|^2+|N_2|^2)}\times\sum_{k_z}(2\cosh \frac{\beta E_1^+}{2})^2 (2\cosh \frac{\beta E_2^+}{2})^2.
\eea
Free energy $F$ of this system is given by
\bea
F&=&-\frac{1}{\beta}\log Z \nonumber \\
&=&\frac{L\Lambda}{\pi} U(|N_1|^2+|N_2|^2)-\frac{2}{\beta}\sum_{k_z}[\log (2\cosh\frac{\beta E_1^+}{2})+\log (2\cosh\frac{\beta E_2^+}{2})].
\eea
Minimizing $F$ with respect to $|N_i|$ (i=1,2), we obtain the following self-consistent equations:
\be
0=\frac{\partial F}{\partial |N_i|}=2\frac{L\Lambda}{\pi}U|N_i|-\sum_{k_z}\tanh \frac{\beta E_i^+}{2} \frac{U^2|N_i|}{E_i^+}. \nonumber \\
\ee
Notice that the self-consistent equation for each order parameter is decoupled from each other. Therefore, we will discuss only one of the two nematic orders, for example $N_1$.

\section{Finite temperature effect}

In this section, we will be discussing how a finite temperature will affect the appearance of different phases. In general, there should exist a critical temperature $T_c$ that characterizes a finite temperature phase transition from a nematic (or CDW) ordered phase to an unordered gapless phase. At critical temperature $T_c$, order parameter vanishes so that we can perform the integration in the self-consistent equations:
\bea
1&=&\frac{U}{\Lambda}g\int_{-\Lambda}^{\Lambda} \frac{dk_z}{4}\tanh \frac{\beta E_i^+}{2} \frac{1}{E_i^+} \nonumber \\
&=&\frac{Ug}{4\Lambda}\int_{-\Lambda}^{\Lambda}dk_z\tanh \frac{\beta |m(k_z)|}{2} \frac{1}{|m(k_z)|} \nonumber \\
&=&\frac{Ug}{2v_f\Lambda}\int_{0}^{v_f\Lambda} d m(k_z)\tanh \frac{\beta |m(k_z)|}{2} \frac{1}{|m(k_z)|} \nonumber \\
&=&\frac{Ug}{v_f\Lambda}\int_{0}^{\beta v_f\Lambda/2} d x \frac{\tanh x}{x} \nonumber \\
&=&\frac{Ug}{v_f\Lambda}[(\tanh x \log x) |_{0}^{\beta v_f{\Lambda}/2}-\int_{0}^{\beta v_f\Lambda/2} d x \frac{\log x}{\cosh^2 x}] \nonumber \\
&\approx&\frac{Ug}{v_f\Lambda}[\log \frac{\beta v_f{\Lambda}}{2}-\int_{0}^{\infty} d x \log x\frac{\log x}{\cosh^2 x}] \nonumber \\
&=&\frac{Ug}{v_f\Lambda}[\log \frac{\beta v_f{\Lambda}}{2}-\log \frac{4e^{\gamma}}{\pi}] \nonumber \\
&=&\frac{Ug}{v_f\Lambda}\log \frac{4e^{\gamma}v_f{\Lambda}}{2\pi k_B T_c}, \nonumber \\
&&
\eea
where $\gamma=0.577...$ is the Euler constant and $N(0)$ is the density of states in 1D. In the integration measure, we have considered the Landau level degeneracy in the $x$-$y$ plane:
\be
g=\frac{S}{2\pi l^2_B}=\frac{eBS}{h}
\ee
Here, $S$ is the surface area of a Dirac semimetal sample spanned in the $x$-$y$ plane. We also take the low temperature limit $T\rightarrow 0$, so that $\beta\omega_{\Lambda}\rightarrow \infty$. Therefore, we arrive at the relation between critical temperature $T_c$ and interaction strength $U$,
\be
T_c=\frac{2e^{\gamma} v_f{\Lambda}}{\pi k_B}e^{-\frac{v_f\Lambda}{Ug}}=\frac{2e^{\gamma}v_f{\Lambda}}{\pi k_B}e^{-\frac{v_f\Lambda}{U}\frac{h}{eBS}}.
\ee
In this expression, we could clearly see that a larger $U$ will naturally lead to a higher $T_c$. Interaction strength, however, is usually determined by the intrinsic properties of a material, and can barely be changed. Instead, we can increase the magnitude of the applied magnetic field which will enhance the transition temperature in a similar way. A simple estimation can be made for $T_c$: if we take the sample in-plane area $S=1\mu m^2$, magnetic field $B=1$ T, interaction strength $U=0.001$ eV, then $T_c$ turns out to be $1000$ K. However, if sample area $S$ is decreased to $0.5\mu m^2$, $T_c=210$ K. If sample area $S$ is further decreased to $0.2\mu m^2$, $T_c=1.8$ K. Decreasing sample area is equivalent to decreasing magnetic field, since both quantities will influence Landau level degeneracy in the same way. This strong scaling behavior reflects the essential role of Landau level degeneracy in our discussions. Therefore, to observe the ordered phase (either nematic phase or CDW phase) we proposed, it is very important to prepare a sample of good enough quality and apply strong enough magnetic field.

\section{Zero temperature limit and gap scaling}
\begin{figure}[t]
\includegraphics[width=0.65\textwidth]{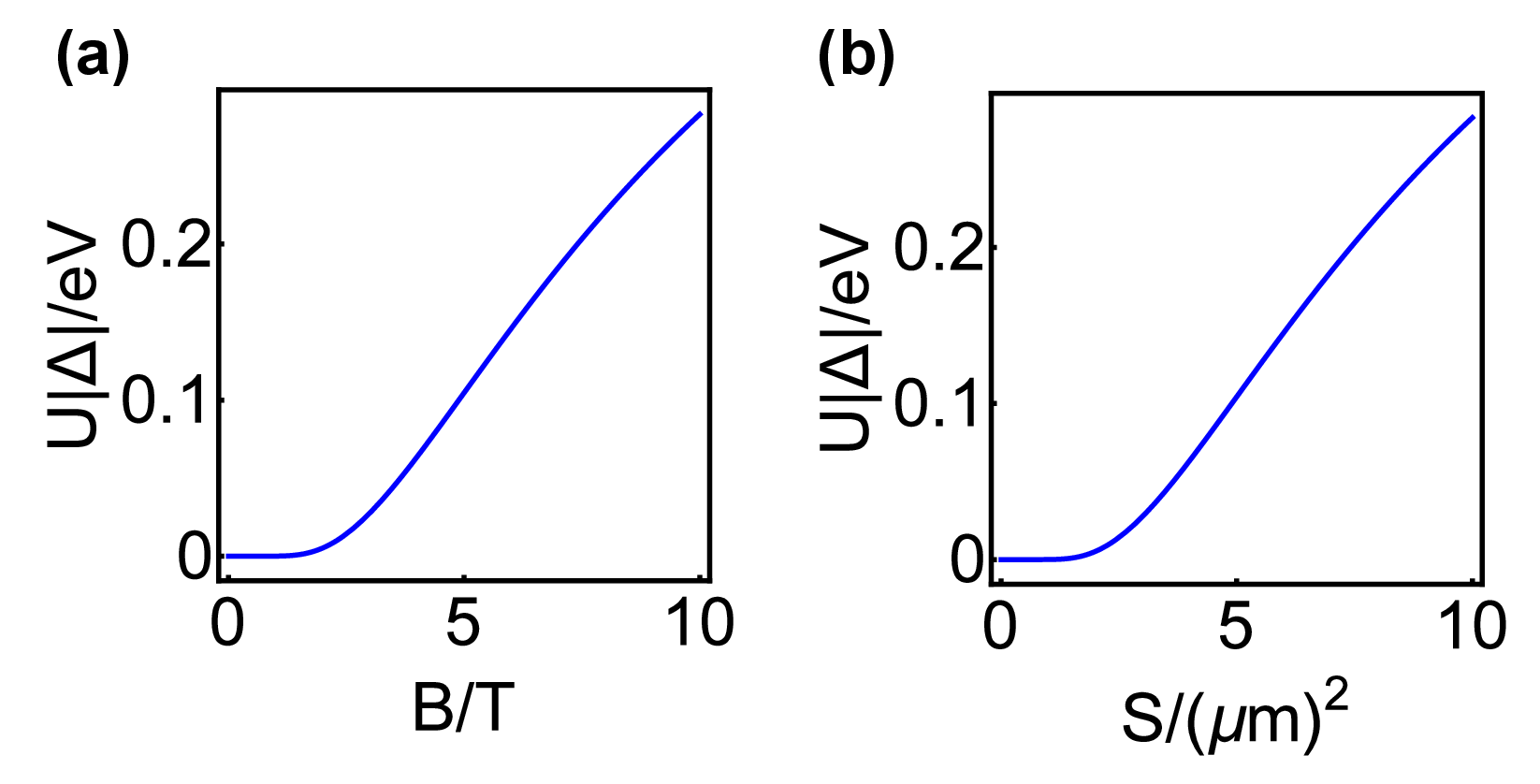}
\caption{Scaling of order parameter $U|\Delta|$ (energy gap) are shown in: (a) Area $S=1\mu m^2$ and (b) $B=1T$. Here, we have adopted the parameters from Ref. \cite{wang2012} and obtained $v_f\approx 1.9$ eV$\cdot${\mbox{\normalfont\AA}}. A momentum cut-off $\Lambda=0.2$ {\mbox{\normalfont\AA}}$^{-1}$ is applied for the calculations. }
\label{Figure:delta-scaling}
\end{figure}

Next, let us look at the zero temperature limit. The free energy can be simplified to
\be
F=\frac{L\Lambda}{\pi}U(|N_1|^2+|N_2|^2)-\sum_{\bf k}(E_1^+ + E_2^+).
\ee
Since $|N_1|$ and $|N_2|$ are decoupled in the self-consistency equations, for $|N_1|$ the self-consistency equation is given by
\bea
&&\frac{\partial F}{\partial |N_1|}=0 \nonumber \\
&\Longleftrightarrow& \frac{2L\Lambda}{\pi}U|N_1|=Lg\int\frac{d k_z}{2\pi}\frac{U^2|N_1|}{\sqrt{m(k_z)^2+U^2|N_1|^2}} \nonumber \\
&\Longleftrightarrow& \frac{1}{Ug}=\frac{1}{4\Lambda}\int_{-\Lambda}^{\Lambda} d k_z \frac{1}{\sqrt{m(k_z)^2+U^2|N_1|^2}}
=\frac{1}{2\pi v_f\Lambda}\log[\frac{2v_f\Lambda}{U|N_1|}].
\eea
Here, $\Lambda$ is the momentum cut-off and we define the Fermi velocity as $v_f=|\frac{m(k)}{k_z}|$. Then, the interaction-induced energy gap is \cite{shankar1994}:
\be
|N_1|\approx\frac{2v_f\Lambda}{U}e^{-\frac{2\pi v_f\Lambda}{Ug}}=\frac{2v_f\Lambda}{U}e^{-\frac{2\pi v_f\Lambda}{U}\frac{h}{eBS}}.
\label{Eq:order}
\ee
Therefore, for an arbitrarily small $U$, a non-zero order (gap) will be developed.

Based on Eq. \eqref{Eq:order}, we are able to check the scaling relation of order parameter (gap) in terms of magnetic field $B$ and sample area $S$. Numerically, these scaling relations are shown in Fig. \ref{Figure:delta-scaling}: (a) We keep area $S=1\mu m^2$ and change the magnetic field $B$. (2) We keep $B=1$ T and change the area $S$. Since Landau level degeneracy $g\approx S\times B$, increasing either $B$ or $S$ will both increase the magnitude of interaction induced gap $|\Delta|$.

\section{Density of states (DOS) and possible experimental detection}
To study possible interaction effect in a rotational symmetry protected 3D Dirac semimetal, we have proposed in the main text to visualize charge distribution by performing local density of states (LDOS) measurement with an STM setup. The appearance of an anisotropic charge distribution is identified as a key feature of the nematic phase. From a different perspective, the development of nonzero ordering also results in a finite gap in the energy spectrum. The energy gap of a system, however, is always ready to be read directly from the DOS measurement near the Fermi level, with the help of an STM setup. Therefore, in this section, we will discuss in details about the DOS feature of a Dirac semimetal sample placed in a strong magnetic field, where magnetic catalysis will assist the formation of ordered states.

First of all, we would like to point out that a DOS measurement (or equivalently gap measurement) is only a direct evidence of the formation of a gap (symmetry breaking). However, such DOS measurement cannot distinguish a nematic phase from a charge density wave (CDW). Therefore an additional LDOS measurement is always necessary to determine the patterns of ordering before any conclusion can be reached.

\begin{figure}[t]
\includegraphics[width=0.65\textwidth]{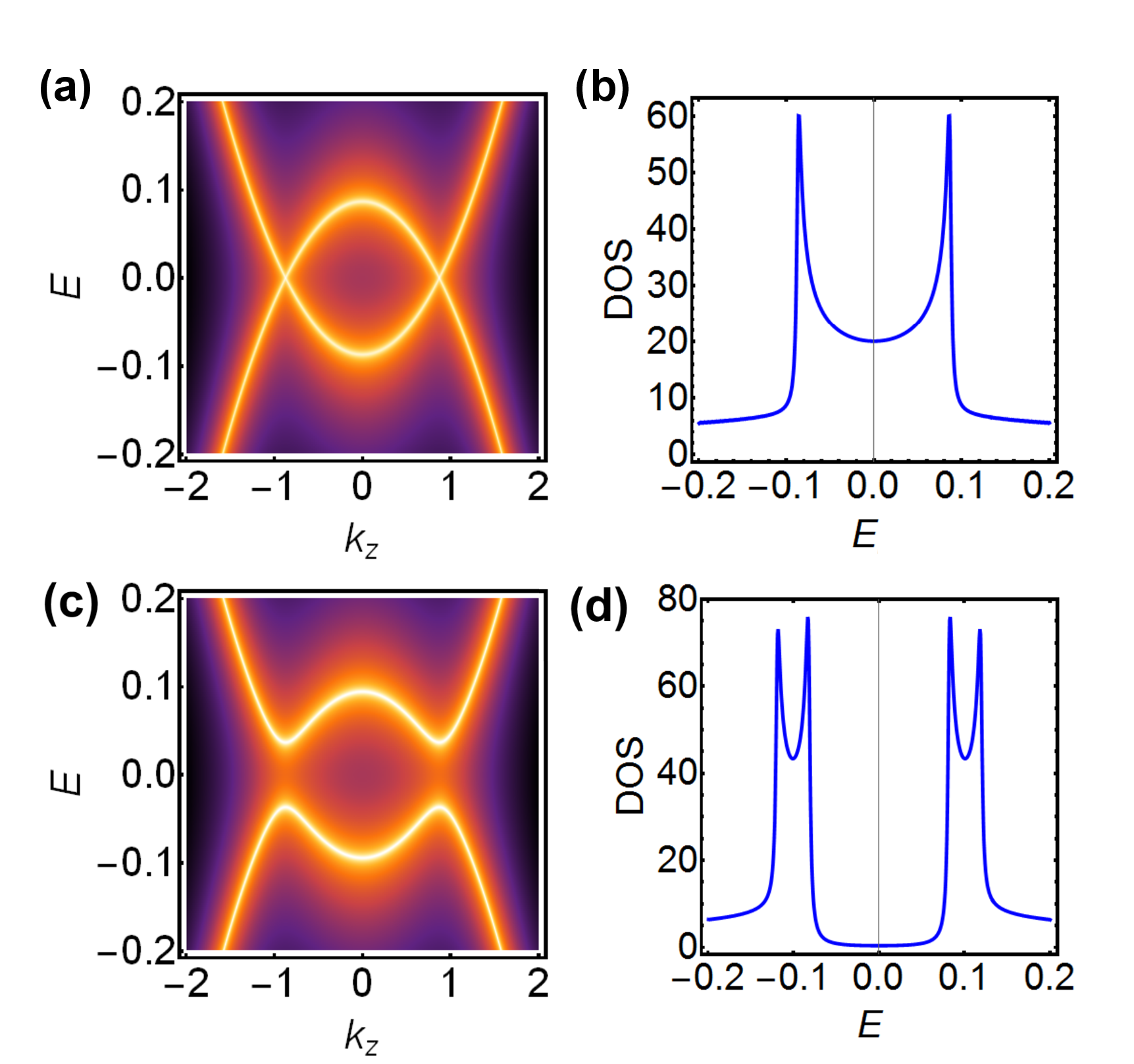}
\caption{Dispersion of zeroth Landau levels are plotted for: (a) $B=1$T, $|\Delta|=0.0$ eV. (c) $B=1$T, $|\Delta|=0.1$ eV. The corresponding DOS plots are shown in (b) and (d). All parameters are adopted from Ref. \cite{wang2012}.}
\label{Figure:LLL-DOS}
\end{figure}

To start, we first consider a simplified problem where only the lowest Landau levels (LLLs) are present. The Hamiltonian of two gapless LLLs is $H_0^{LLL}(k_z)=(M_0-M_1k_z^2-\frac{M_2}{l^2})\sigma_z$. Here $\sigma_{x,y,z}$ are Pauli matrices defined under the bases $|\Psi\rangle=(|\frac{1}{2}\rangle,|-\frac{3}{2}\rangle)^T$. In the discussions below, we will focus on the case of nematic phase where translational symmetry is preserved. Then a complex nematic order parameter $\Delta$ can show up in the off-diagonal part of $H_0^{LLL}$,
\begin{equation}
H^{LLL}(k_z)=
\begin{pmatrix}
M_0-M_1k_z^2-\frac{M_2}{l^2} & \Delta \\
\Delta^* & -(M_0-M_1k_z^2-\frac{M_2}{l^2})
\end{pmatrix}.
\end{equation}
Generally, for a one dimensional Hamiltonian $H(k_z)$, the DOS $\rho(E)$ at energy $E$ can be expressed in terms of retarded Green function $G^R(E,k_z)$,
\begin{eqnarray}
&G^R(E,k_z)&=\frac{1}{E-H(k_z)+i\eta}, \nonumber \\
&\rho(E,k_z)&=-\frac{1}{\pi}\text{Im}\{\text{Tr}[G^R(E,k_z)]\}, \nonumber \\
&\rho(E)&=\int \frac{dk_z}{2\pi}\rho(E,k_z)=-\frac{1}{\pi}\text{Im}\int \frac{dk_z}{2\pi}\text{Tr}[G^R(E,k_z)].
\end{eqnarray}

Here. $\eta\ll1$ is a small number to avoid singularity. We have calculated both $\rho(E,k_z)$ and $\rho(E)$ for $H_{k_z}^{LLL}$ and obtained band dispersions as well as the corresponding DOS figure. The DOS $\rho(E)$ has an arbitrary unit because its calculated value is determined by the value of $\eta$ we are choosing, and therefore only the relative magnitude of DOS within the same DOS plot is physically meaningful. As shown in Fig. \ref{Figure:LLL-DOS} (a) and (b), when $|\Delta|=0$, the system is gapless and the DOS of the 1D Dirac point ($E=0$) is finite. Notice that in Fig. \ref{Figure:LLL-DOS} (b), the DOS is diverging (peaks of DOS) at two different $E$, which corresponds to two band extreme around $E=\pm 0.1$. When $|\Delta|=0.1$eV is turned on, the system is gapped (Fig. \ref{Figure:LLL-DOS} (c)) and the DOS within the energy gap is suppressed. A new band edge formed around the energy $E=-0.003$ and $E=0.009$, leading to two additional DOS peaks.

\begin{figure}[t]
\includegraphics[width=0.9\textwidth]{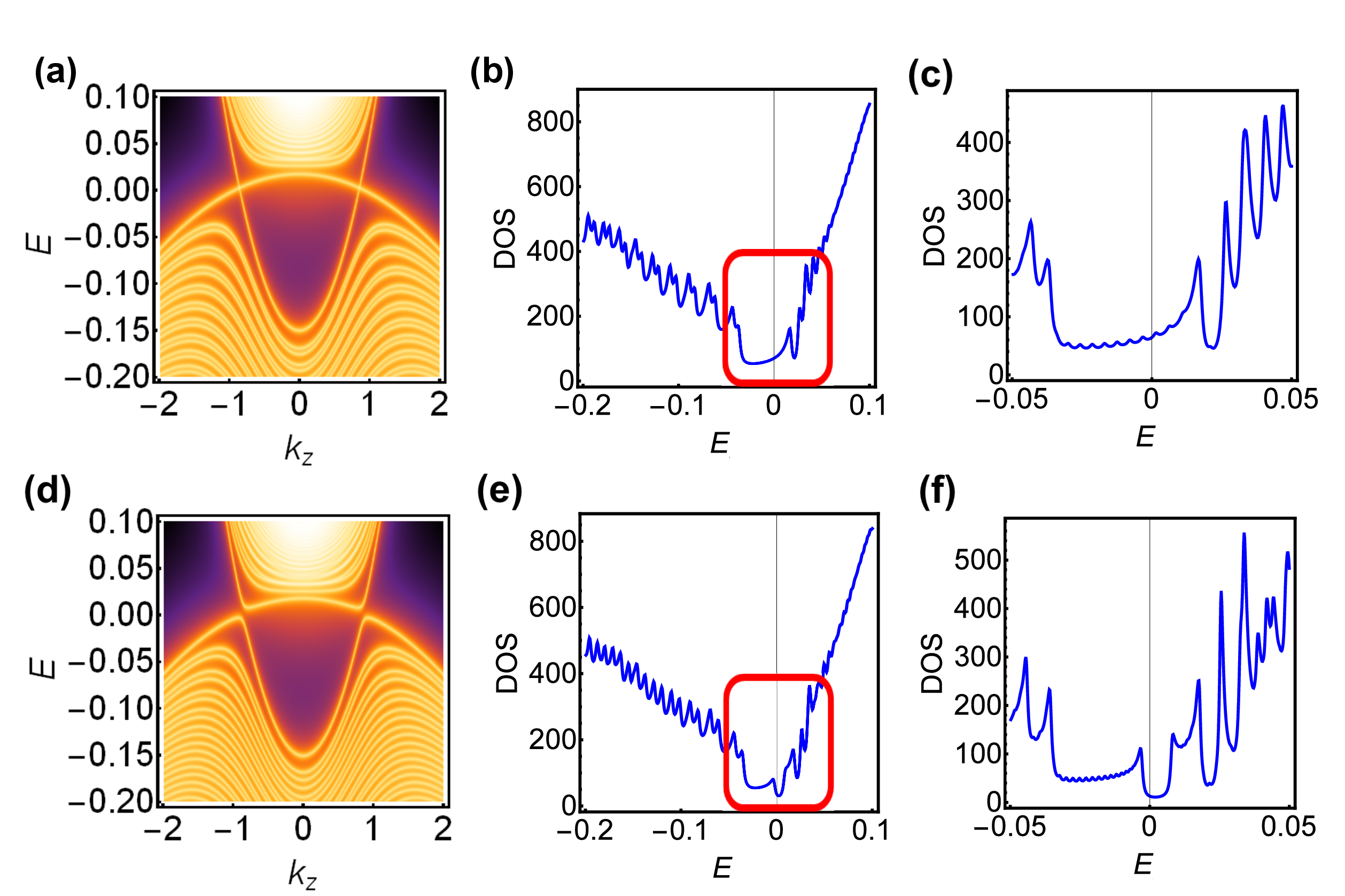}
\caption{Dispersion of Landau levels in Dirac semi-metals are plotted for: (a) $B=20$T, $|\Delta|=0.0$ eV. (d) $B=20$T, $|\Delta|=0.01$ eV. The corresponding DOS plots are shown in (b) and (e). In (c) and (f), we zoom in to the red block region to get a better view of the DOS around $E=0$.}
\label{Figure:LL-DOS}
\end{figure}

Next, we consider a more realistic model where higher Landau levels are present (See Eq. \eqref{Eq:Landau level} for details). As shown in Fig. \ref{Figure:LL-DOS} (b) and (e), if we tune $E$ continuously, a peak of DOS will show up when $E$ coincides with the band extreme of a Landau level. If we focus only on the low energy DOS around $E=0$, as shown in the red block regions in (b) and (e), the DOS plots in (c) and (f) capture the main features of earlier discussions in Fig. \ref{Figure:LLL-DOS}. Experimentally, Fig. \ref{Figure:LL-DOS} (f) will be a smoking-gun signature of interaction induced ordering in rotational symmetry protected Dirac semi-metals.

Unfortunately, the existing experiments are not intended for finding the nematic phase, although all the necessary conditions should already exist. The most closely related experiment is the STM measurement of Cd$_2$As$_3$ from Yazdani's group \cite{jeon2014}. They even include a discussion of Landau level spectrum for magnetic fields along different directions (Fig. 4d and e in \cite{jeon2014}), showing that two zero Landau levels will cross each other for magnetic field along $[001]$ direction and anti-cross each other for $[112]$-directional magnetic field. Our prediction is that even for $[001]$-directional magnetic field, one still finds an anti-crossing behavior due to interaction effect. However, the STM measurement in Yazdani's experiment is implemented on the $[112]$ surface, which breaks $C_4$ rotation by itself. This prevents the observation of the nematic phase. To search for nematic phases, an STM measurement along the $[001]$ surface is required.

Another related experiment is the quantum oscillation of magneto-transport measurement in Cd$_2$As$_3$ \cite{cao2014}. Landau level splitting is resolved by rotating magnetic fields in the quantum oscillation measurements. However, this experiment can only reach the Landau level $N\geq2$. Thus, to observe our prediction, one needs to further lower electron density to reach the truly quantum limit with experimentally feasible magnetic fields. In Ref. \cite{zhao2015}, the quantum limit is reached at around $43$T. However, the magnetic field is applied along the [112] direction, which again breaks the $C_4$ rotation symmetry.

\end{document}